\documentclass{article}

\usepackage{PRIMEarxiv}
\usepackage{amsmath}
\usepackage[utf8]{inputenc} % allow utf-8 input
\usepackage[T1]{fontenc}    % use 8-bit T1 fonts
\usepackage{hyperref}       % hyperlinks
\usepackage{url}            % simple URL typesetting
\usepackage{booktabs}       % professional-quality tables
\usepackage{amsfonts}       % blackboard math symbols
\usepackage{nicefrac}       % compact symbols for 1/2, etc.
\usepackage{microtype}      % microtypography
\usepackage{lipsum}
\usepackage{graphicx}
\graphicspath{{media/}}     % organize your images and other figures under media/ folder
\title{Transcranial Photoacoustic Imaging for Human Intracranial Pressure Evaluation}

\author{
Ruixi Sun\textsuperscript{1,2,3,\#}\quad
Hengrong Lan\textsuperscript{1,2,\#}\quad
Yuanyuan Dang\textsuperscript{4,\#}\\
Yunhui Jiang\textsuperscript{1,2}\quad
Youshen Xiao\textsuperscript{1,2,3}\quad
Sheng Liao\textsuperscript{1,2}\quad
Fan Zhang\textsuperscript{1,2}\\
Daohuai Jiang\textsuperscript{5}\quad
Zhongqi Li\textsuperscript{1,2,*}\quad
Hulin Zhao\textsuperscript{4,*}\quad
Fei Gao\textsuperscript{1,2,6,*}\\[6pt]
\textsuperscript{1}\,School of Biomedical Engineering, Division of Life Sciences and Medicine,\\
University of Science and Technology of China, Hefei, Anhui, 230026, China\\
\textsuperscript{2}\,Hybrid Imaging System Laboratory, Suzhou Institute for Advanced Research,\\
University of Science and Technology of China, Suzhou, Jiangsu, 215123, China\\
\textsuperscript{3}\,School of Information Science and Technology, ShanghaiTech University, Shanghai 201210, China\\
\textsuperscript{4}\,Department of Neurosurgery, Chinese PLA General Hospital, Beijing, 100853, China\\
\textsuperscript{5}\,College of Photonic and Electronic Engineering, Fujian Normal University, Fuzhou, Fujian, China\\
\textsuperscript{6}\,School of Engineering Science,\\
University of Science and Technology of China, Hefei, Anhui, 230026, China\\[6pt]
\textbf{Correspondence:} \texttt{jasonlee0091@outlook.com, zhaohulin\_90@sohu.com, fgao@ustc.edu.cn}\\[4pt]
\textsuperscript{\#}\,Co-first authors \quad \textsuperscript{*}\,Corresponding authors
}

\begin{document}
\maketitle

\begin{abstract}
Photoacoustic imaging (PAI), by combining high optical contrast with ultrasonic resolution, offers a promising noninvasive approach for dynamic monitoring of cerebral vasculature. However, transcranial PAI still faces significant challenges due to strong attenuation of both optical and acoustic signals by the skull. In this study, we propose a multi-wavelength photoacoustic tomography system and method for intracranial pressure (ICP) assessment, enabling visualization of cross-sectional structures of the middle cerebral artery (MCA) through the human temporal bone. By utilizing multi-wavelength excitation in the near-infrared-I (NIR-I) window, quantitative maps of blood oxygen saturation ($\mathbf{sO_2}$) are reconstructed, and the relationship between oxygenation dynamics and ICP variations is established. Experimental results demonstrate that the proposed system can successfully capture dynamic $\mathbf{sO_2}$ fluctuations in the MCA despite skull attenuation, revealing its characteristic responses to ICP changes. This work provides a high-precision, noninvasive imaging tool for early stroke diagnosis, cerebral vascular function assessment, and neurointerventional guidance, highlighting the clinical translational potential of PAI in neuroscience.
\end{abstract}

% keywords can be removed

\section{Introduction}
Accurate diagnosis and functional evaluation of cerebrovascular diseases are critical for the prevention and treatment of major neurological disorders such as stroke, aneurysms, and intracranial pressure (ICP) abnormalities\cite{kim2025non,prust2024addressing,ren2021nir}. As one of the primary arteries supplying blood to the brain, the middle cerebral artery (MCA) plays a key role in cerebral circulation, and its morphological abnormalities, hemodynamic disturbances, and microenvironmental changes associated with ICP fluctuations are closely linked to pathological processes such as ischemic stroke, vascular dementia, and elevated ICP \cite{klijn2010haemodynamic,claassen2021regulation}. However, current clinical imaging modalities such as magnetic resonance angiography (MRA)\cite{logothetis2008we} and computed tomography angiography (CTA)\cite{thilak2024diagnosis} suffer from limitations including low temporal resolution (on the order of seconds to minutes)\cite{prevedel2025three} and ionizing radiation exposure, making them unsuitable for high-sensitivity, noninvasive monitoring of dynamic blood flow and ICP variations. Optical microscopy techniques, such as 
three-photon microscopy\cite{prevedel2025three}, offer micron-level resolution but are limited to superficial vasculature due to the strong light scattering by the skull. Some recent studies have attempted imaging through the temporal bone to mitigate skull-induced signal loss. Transcranial ultrasound imaging, for instance, uses the temporal bone as an acoustic window to noninvasively assess blood flow in major cerebral arteries. Yet it relies on round-trip acoustic propagation, making it highly susceptible to signal attenuation and distortion caused by the skull, especially in elderly or calcified individuals. Therefore, the development of a novel vascular imaging modality that simultaneously offers deep tissue penetration, high spatiotemporal resolution, and multimodal functionality (including blood oxygenation and ICP monitoring) has become a pressing challenge in the field of neuroimaging\cite{mozaffarzadeh2022refraction}.

Photoacoustic imaging (PAI) provides an innovative solution to this problem by combining optical excitation with ultrasonic detection. Leveraging the photoacoustic (PA) effect of endogenous chromophores such as hemoglobin, PAI enables high-contrast imaging of vascular structures at millimeter to centimeter depths, and further allows dynamic monitoring of blood oxygen saturation ($\mathbf{sO_2}$) through multi-wavelength excitation\cite{jiang2024photoacoustic,tian2024image}. While PAI has demonstrated great potential in in vivo imaging of cortical vasculature, transcranial PAI still faces significant technical bottlenecks: the skull causes strong absorption and scattering of PA signals, reducing both image resolution and signal-to-noise ratio, and deep-seated major vessels such as the MCA are often difficult to detect due to complex hemodynamic environments.

Previous studies have made preliminary attempts at transcranial PAI of the human brain. For example, Zhang et al. \cite{zhang2022transcranial} performed functional PA tomography (fPAT) on a patient who had undergone hemicraniectomy, demonstrating the imaging potential of fPAT in the absence of a skull barrier. Ni et al. \cite{ni2022noninvasive} achieved superficial vascular imaging through the temporal bone in healthy volunteers and validated their structural results using time-of-flight MRA. However, the former requires invasive craniotomy and is not clinically generalizable, while the latter, though noninvasive, is limited in imaging depth and does not address functional assessment related to ICP.

To improve the performance of transcranial PAI, we propose a novel MCA imaging approach based on linear-array detection. By scanning over the temporal window, this method reduces skull-induced PA signal attenuation. It further incorporates multi-wavelength spectral imaging to extract blood oxygenation parameters. Using this system, we successfully visualized deep intracranial arteries beneath the temporal bone in multiple human subjects, with structural validation performed via magnetic resonance imaging (MRI), confirming anatomical consistency. In summary, this study demonstrates that transcranial PAI through the temporal window enables noninvasive estimation of ICP abnormalities, offering a powerful tool for understanding cerebral microcirculation regulation mechanisms and laying the foundation for future clinical applications in diagnosis and treatment of brain diseases.

\begin{figure}[h]
\centering
\includegraphics[width=3in]{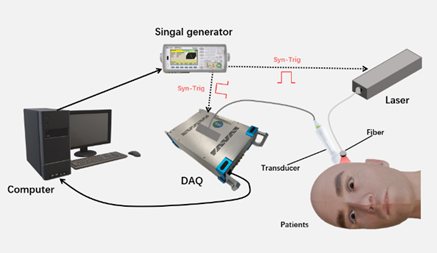}
\caption{Schematic of the Experimental Setup.}
\label{fig_1}
\end{figure}

\section{Method}
\subsection{Transcranial Photoacoustic Imaging System}

The transcranial PAT system used in this study is illustrated in Fig. 1. In PAI mode, the system employs 21 excitation wavelengths ranging from 700 to 900 nm in 20 nm increments to enable multispectral blood oxygenation acquisition. The laser excitation is provided by a tunable OPO (model: pai-NIR1, TsingPAI Technology Pte Ltd., China), featuring high output stability and programmable wavelength control. The pulse repetition frequency (PRF) is set to 10 Hz, with automatic wavelength switching between successive pulses to achieve rapid full-spectrum scanning. To ensure energy consistency, the pulse energy at each wavelength is calibrated using a power meter prior to imaging. The maximum pulse energy at the skin surface is 30 mJ, well below the safety threshold.

PA signals are detected using a one-dimensional linear-array ultrasound transducer (128-element, Doppler Inc., China), consisting of 128 piezoelectric elements with a center frequency of 7.5 MHz and a bandwidth of 73\%. The signals are digitized using a high-speed data acquisition card (model: HISonics, HIS PATech Pte. Ltd.) at a sampling rate of 40 MSPS (mega samples per second).

The transducer is fixed over the temporal window using a custom-designed 3D-printed holder. The diffused laser beam is directed onto the same region and coupled to the skin via ultrasound gel to optimize acoustic transmission. The scanning is performed on the skin surface over the left temporal bone, and all subjects are required to wear laser safety goggles to avoid potential exposure.

\subsection{Image Reconstruction and Quantitative Blood Oxygen Algorithms}

The optical absorption of deoxygenated hemoglobin $(\mathrm{HbR})$ is significantly higher than that of oxygenated hemoglobin $\left(\mathrm{HbO}_2\right)$ in the $700-800 \mathrm{~nm}$ wavelength range\cite{yao2014sensitivity}. Conversely, in the $800-900 \mathrm{~nm}$ range, $\mathrm{HbO}_2$ exhibits greater absorption than HbR. Considering that tissue scattering increases substantially below 700 nm , potentially introducing significant errors, we selected the 700-900 nm range for quantitative PAI.

The algorithm used in this study is based on linear spectral fitting to calculate $\mathbf{sO_2}$\cite{li2008simultaneous}.

\[
P(\lambda_i, x, y) = \Phi(\lambda_i) \cdot \left[
\begin{aligned}
&\varepsilon_{\mathrm{HbR}}(\lambda_i)\, C_{\mathrm{HbR}}(x, y) \\
+{} &\varepsilon_{\mathrm{HbO}_2}(\lambda_i)\, C_{\mathrm{HbO}_2}(x, y)
\end{aligned}
\right]
\]

Here, $\Phi\left(\lambda_i\right)$ represents the local optical fluence, and $P\left(\lambda_i, x, y\right)$ denotes the reconstructed PA image at wavelength $\lambda_i$. The terms $\varepsilon_{\mathrm{HbR}}\left(\lambda_i\right)$ and $\varepsilon_{\mathrm{HbO}_2}\left(\lambda_i\right)$ are the molar extinction coefficients of HbR and $\mathrm{HbO}_2$ at wavelength $\lambda_i$, while $C_{\mathrm{HbR}}(x, y)$ and $C_{\mathrm{HbO}_2}(x, y)$ represent their corresponding molar concentrations at position $(x, y)$.

After fluence normalization, the concentrations $C_{\mathrm{HbR}}(x, y)$ and $C_{\mathrm{HbO}_2}(x, y)$ can be calculated by solving the following linear system using multiple wavelength measurements:

$$
\left[\begin{array}{c}
C_{\mathrm{HbR}}(x, y) \\
C_{\mathrm{HbO} O_2}(x, y)
\end{array}\right]=\left(\varepsilon^T \varepsilon\right)^{-1} \varepsilon^T P
$$

where

$$
P=\left[\begin{array}{l}
P\left(\lambda_1, x, y\right) \\
P\left(\lambda_2, x, y\right)
\end{array}\right]
$$

is the PA measurement vector at two wavelengths, and

$$
\varepsilon=\left[\begin{array}{ll}
\varepsilon_{\mathrm{HbR}}\left(\lambda_1\right) & \varepsilon_{\mathrm{HbO}_2}\left(\lambda_1\right) \\
\varepsilon_{\mathrm{HbR}}\left(\lambda_2\right) & \varepsilon_{\mathrm{HbO}_2}\left(\lambda_2\right)
\end{array}\right]
$$
is the extinction coefficient matrix.

PA image reconstruction was performed using a delay-and-sum (DAS) algorithm. As photon scattering in tissue was not corrected during fluence estimation, the resulting oxygen saturation is defined as a "PA oxygenation index", which reflects relative changes in intracranial blood oxygenation rather than absolute $\mathrm{sO}_2$ levels.

\subsection{Overview of the Data Processing Workflow}

The proposed data processing pipeline converts raw PA signals into blood $\mathbf{sO_2}$ distribution maps. First, the system uses laser trigger signals to establish correct correspondence between each A-line and its associated excitation wavelength, ensuring accurate spectral labeling of the multi-wavelength PA data during subsequent processing. Next, background removal is applied to the raw 
PA images by thresholding to eliminate low-SNR signals from non-vascular regions. Specifically, the threshold is set to five times the standard deviation (STD) of the system noise, and pixels below this threshold are excluded from the spectral unmixing process.

Subsequently, the signals are normalized by the incident pulse energy to compensate for laser energy fluctuations across pulses and wavelengths. Assuming the beam profiles of the two wavelengths are identical, the local optical fluence at the skin surface is considered proportional to the laser pulse energy. Therefore, the acquired signals can be normalized using this value.

\section{Experiments \& Results}
\subsection{Participants and Ethical Approval}
A total of nine patients with intracranial lesions undergoing PAI combined with invasive ICP monitoring were enrolled in this study. All participants provided informed consent prior to inclusion. All subjects met the clinical indications for ICP monitoring.

Table 1 summarizes the baseline characteristics of the enrolled patients, including ICP  measurements ($\mathbf{mmH_2O}$) and the corresponding PA oxygenation indices.
\begin{table*}[htbp]
\centering
\caption{ICP and Oxygenation Parameters in Patients}
\label{tab:icp_oxygen_full}
\renewcommand{\arraystretch}{1.2}
\setlength{\tabcolsep}{6pt}
\begin{tabular}{lccccccccc}
\toprule
\textbf{Patient ID}        & 1     & 2     & 3     & 4     & 5     & 6     & 7     & 8     & 9     \\
\midrule
ICP($\mathbf{mmH_2O}$)                 & 80    & 140   & 150   & 330   & 270   & 160   & 40    & 170   & 70    \\
Oxygenation Index         & 0.5986 & 0.5881 & 0.5814 & 0.5789 & 0.5813 & 0.5815 & 0.5954 & 0.5862 & 0.5921 \\
\bottomrule
\end{tabular}
\end{table*}

\begin{figure}[h]
\centering
\includegraphics[width=3in]{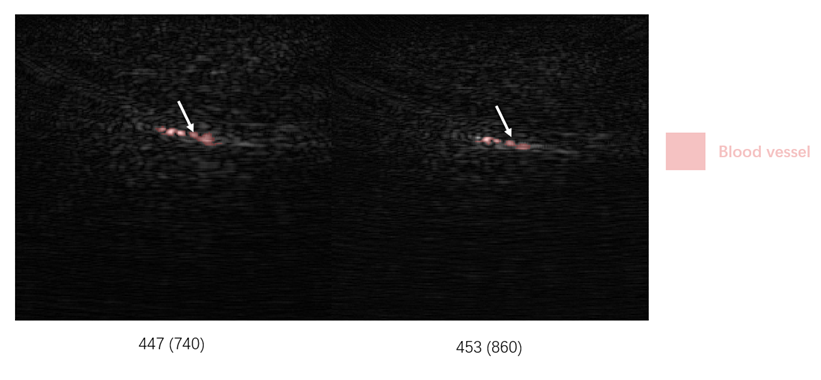}
\caption{Vessel Segmentation under Different Wavelengths.}
\label{fig_1}
\end{figure}
\subsection{Vascular Segmentation Results}
As shown in Fig. 2, the vascular structures in the PA images were successfully segmented. The results preserve the spatial continuity of major vessel trunks and demonstrate the ability to identify fine branches even in high-noise regions (indicated by arrows).
\begin{figure}[h]
\centering
\includegraphics[width=3in]{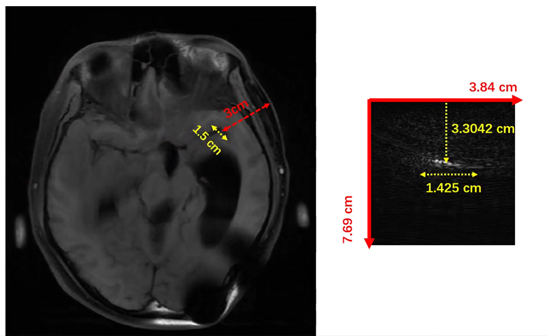}
\caption{Comparison of MRI and PA Cerebral Artery Depths.}
\label{fig_1}
\end{figure}
\subsection{Registration with MRI}
In addition, we further validated that the signals acquired by PAI originated from the MCA by matching them with MRI-derived anatomical features. As shown in Fig. 3, we first performed quantitative measurements of key geometric parameters of the MCA—namely, its depth and width—based on the MRI data.
\begin{figure}[h]
\centering
\includegraphics[width=3in]{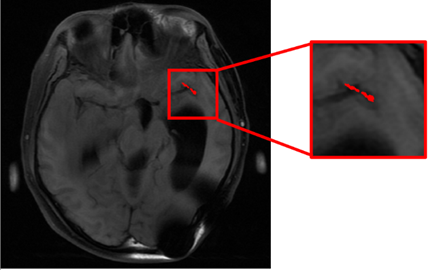}
\caption{Registration of MRI and PA Cerebral Arteries.}
\label{fig_1}
\end{figure}
Subsequently, the same parameter extraction strategy was applied to the PA images to obtain the depth and width of the MCA. A comparison between the two datasets revealed a high degree of consistency in both spatial scale and morphological characteristics. As shown in Fig. 4, the segmented vessel regions from PAI closely correspond to the MCA identified in the registered MRI images, indicating strong agreement between the two modalities.

\section{Conclusion}
This study presents a novel method for noninvasive brain oxygenation monitoring based on PAI. Experimental results confirm that this method enables the acquisition of MCA oxygenation information under noninvasive conditions and provides a potential biomarker for ICP variation. Although technical challenges remain, the findings lay an important foundation for clinical translation of PAI in neuroscience. Future integration with multimodal imaging and intelligent algorithms is expected to further enhance the clinical value of this technology.

%Bibliography
\bibliographystyle{unsrt}  
\bibliography{references}

\end{document}